\documentclass[useAMS,usenatbib]{mn2e}

\usepackage{natbib}
\usepackage{amsmath,amsfonts,amssymb}
\usepackage{epsfig}
\usepackage{ifthen}
\usepackage{color}
\usepackage{xspace}

\def\pr{{\rm Pr}}
\def\data{\boldsymbol{d}}
\def\mass{\Sigma}
\def\pars{\mathbf{x}}

\def\pdf{PDF\xspace}

\def\eg{{\em e.g.}\xspace}
\def\ie{{\em i.e.}\xspace}
\def\citeLE{M02\xspace}
\def\LE{{\scshape LensEnt2}\xspace}
\def\Natoms{n\xspace}

\title[Physical component analysis]%
{Physical component analysis of\\
galaxy cluster weak gravitational lensing data}
\label{firstpage}
\author[Marshall]{
Phil Marshall\\
Kavli Institute for Particle Astrophysics and Cosmology, Stanford University
PO Box 20450, MS29, Stanford, CA 94309, USA\\
}
\date{Accepted 2006 August 10. Received 2006 August 10; in original form 2005 November 9.}
\pagerange{\pageref{firstpage}--\pageref{lastpage}}
\pubyear{2006}

\begin{document}
\maketitle


\begin{abstract}

We present a novel approach for reconstructing the projected mass distribution
of clusters of galaxies from sparse and noisy weak gravitational lensing shear
data. The reconstructions are regularised using knowledge gained from numerical
simulations of clusters: trial mass distributions are constructed from $\Natoms$
physically-motivated components, each of which has the universal density 
profile and characteristic geometry observed in simulated clusters. 
The parameters of these components are assumed to be distributed \emph{a priori}
in the same way as they are in the simulated clusters. Sampling mass
distributions from the components' parameters' posterior probability density
function allows estimates of the mass distribution to be generated, with error
bars. 
The appropriate number of components is inferred from the data itself via the
Bayesian evidence, and is typically found to be small, reflecting the quality of
the simulated data used in this work. 
Ensemble average mass maps are found to be robust to the details of the noise
realisation, and succeed in recovering the input mass distribution (from a
realistic simulated cluster) over a wide range of scales.  We comment on the
residuals of the reconstruction and their implications, and discuss the
extension of the method to include strong lensing  information.

\end{abstract}



\section{Introduction}

Mapping the mass distributions of galaxy clusters via their weak gravitational
lensing effect has become something of a standard tool in astrophysics, allowing
these most massive objects to be better understood in terms of their matter
content,  dynamical state and their value as galaxy evolution laboratories and
cosmic observatories. Given this importance, it seems worthwhile to investigate
\ more accurate, more robust, and more practically useful methods for
reconstructing the mass distributions in clusters from the available data.

Following a number of seminal papers on the subject in the
1990s~\citep[\eg][]{GL/TVW90,GL/K+S93}, the emphasis now is very much on the
application of mapping methods to weak gravitational lensing shear data. Large
CCD mosaic cameras such as SuprimeCam at Subaru, MegaCam at CFHT and the ESO
Wide Field Imager at La Silla have enabled the mass distributions of clusters to
be mapped to much larger radii than  before~\citep[\eg][]{GL/C+S02,GL/Bro++05}.
From space with HST, the same science has been made possible in higher redshift
clusters, first though large multi-pointing WFPC2
datasets~\citep[\eg][]{GL/HFK00,GL/Kne++03} and subsequently with observations
with the Advanced Camera for Surveys 
(ACS)~\citep[\eg][]{GL/Lom++05,GL/Jee++05}. As is always the case in astronomy,
these data are being pushed to their limits: the aim now is to understand
cluster mass distributions in great detail, moving beyond the simple mass
estimates of the early years. \citet{GL/Kne++03} and \citet{GL/Gav++03} measured
the outer logarithmic slope of the density profile in two systems, while
\citet{GL/CGM04} investigated the relative peak positions of the gravitating
mass density and the intracluster gas density (the latter being derived from the
X-ray surface brightness). The quantification of the substructure in galaxy
clusters is a topic of ongoing research, with progress being made in the central
parts of clusters by comparing strong and weak lensing mass models with
predictions from N-body simulations~\citep{GL/N+S04}.

Despite the  advances in data quality, weak gravitational lensing data remains
very sparse and very noisy.  It is notable indeed that the most exciting results
in the field in recent years have come from the comparison of weak lensing data
with external observations, such as the modelling of strong gravitational
lensing features \citep[\eg][]{GL/Kne++03,GL/Bra++05b}, the X-ray emission  
\citep[as in][]{GL/CGM04}, and the optical data on the cluster member galaxies
\citep[\eg][]{O/Czo++02,GL/Kne++03}. With this in mind, we ask how best to
extract as much information as possible from the weak lensing data, and how to
draw the most meaningful conclusions about the mass distributions in clusters.
In general terms, including information from external sources means assigning
appropriate prior probability distributions to whatever set of parameters we are
using: to this end we seek a flexible fitting algorithm that is able to cope
with such constraints and return parameter estimates (with accurate
uncertainties) that reflect all the observational data in hand. Moreover, the
choice of model itself can be made so as to facilitate comparisons with 
independent observations. However, in the first instance, it seems sensible to
postpone combination with other observational data until the lensing signal is
understood.

In all weak lensing reconstruction algorithms to date  the assumed model has
been that of a grid of pixels, whose values (be they surface mass density or
lens potential) comprise the model parameters, an approach often described as
``non-parametric.'' It is not at all clear that a grid of pixels  is the optimal
model for a cluster mass distribution; as shown in~\citet[][ hereafter
M02]{GL/Mar++02}, such a large number of parameters is very often discouraged by
the data quality, leading to over-fitting and potentially  over-interpretation
of the data. In \citeLE, the effective number of parameters was reduced by
including the assumption of the mass pixels being correlated on some
characteristic angular scale ($\sim 1\; $arcmin), an algorithm implemented in
the \LE code.  This assumption led to smoother, less noisy maps, which, by
virtue of the resolution scale being inferred from the data themselves, were
restricted to show only ``believable'' structures with angular scales greater
than the resolution parameter. In this work we seek a more natural basis set of
functions with which to model cluster mass distributions. By correlating pixels
together, cluster-like structures can be more easily modelled: the logical
extension of this idea is to build up a mass distribution from components that
already have cluster-like properties.  This is the principal concept in this
work. 

From N-body simulations we expect the ensemble average mass distribution to be
ellipsoidal with an NFW profile~\citep[\eg][]{CS/NFW97,CS/J+S02}, and clusters
to lie at the high mass end of a hierarchy of structures, each with this same
universal profile. The NFW profile has some support from the data, at  least for
the most massive halos: previous gravitational lensing analyses have found the 
NFW profile mass distribution to provide a somewhat better description of the
data than competing models~\citep[\eg][]{GL/C+S02,GL/Kne++03,GL/Gav++03}, as
have high resolution X-ray studies \citep[\eg][]{X/ASF02}.

If all the mass in clusters of galaxies were distributed exactly in ellipsoidal
NFW-profile halos, then the optimal basis set for the lensing inverse problem 
would be a collection of ellipsoidal NFW profile mass components, shifted and
scaled to match the various mass clumps in the field. The results of the
simulations suggest that clusters do indeed look like this,  and it is this that
motivates our choice of mass model. This basis set allows a continuously
multi-scale mass map to be reconstructed, with the angular resolution reflecting
the local data quality and signal strength, but also the expected density
profile cusps and  slopes.  We anticipate that such a basis set will  be able to
cope much better with the high level of  noise in the data, provided that the
data themselves are used to select the appropriate number of mass components
used in the inference: if the weak shear data only support the inference of a
small number of parameters associated with a small number of mass components,
then we must be able to quantify this statement, and so automatically prevent
the over-fitting that can plague pixel-based methods. 

While it is the NFW component parameters that are to be inferred, it is a
reconstructed projected mass map that best encapsulates our state of knowledge
of the cluster potential. Such maps can be constructed by tabulating the
inferred (shifted and scaled) basis functions onto a grid of pixels. These maps
will have, by design, highly correlated pixel values: the covariance of the
pixel values should also be taken into account  to make the maps quantitatively
useful.

The general ideas introduced above, whilst not previously applied in the field
of weak gravitational lensing, are not new to the inferential science community.
Such ``atomic'' methods were suggested and developed for image reconstruction 
by~\citet{ST/Ski98}, who was motivated to move beyond pixellised models when
analysing spectral and image data for the same reasons as outlined above.
Skilling's ``atoms'' were often very simple in nature, as he sought to
reconstruct complex images with very little prior knowledge.  To have such a
natural choice of ``atom,'' as proposed above for galaxy clusters, is something
of a rare treat. Indeed, we might instead dub the NFW halos  ``physical
components'' to emphasise their highly motivated nature. This also helps to
avoid the possible  confusion that may be induced by the term  ``atomic
inference'' among the readership of this journal. 

However, we should remain open to the idea that the details of the component
properties are best also determined from the data -- how else would we learn
that the numerical simulations are realistic? In this work we demonstrate the
use of NFW halos in modelling weak lensing data: alternative models may not have
such well-defined prior distributions, which puts them at a natural disadvantage
when comparing models. However, if a particular dataset demands a different
profile atom then this can be straightforwardly inferred from the data
\citep{GL/Kne++03}. 

The methodology in this work can be rightly seen as an extension of the mass
modelling of \citet[][ and subsequent works]{GL/Kne++96}. In their approach, one
or two smooth  elliptical mass components were used to model the positions of 
the strongly lensed images, with the parameters of the components optimised, and
the model refined, as more multiple image systems were identified; the weak
lensing data was used (if at all) as a weak constraint on the (primarily strong
lensing) model. Here, we adopt and justify the same modelling philosophy, but
focus on the weak lensing effects of lower mass substructure at larger radii,
increasing the number of free parameters, automating their estimation, exploring
parameter degeneracies and pushing the interpretation of the mass components
beyond that of simply stating a best-fit parameter set. Indeed, our method is
much closer to the ``smooth particle inference'' approach put forward for X-ray
data analysis
by~\citet{X/PMA05}: the differences in this case arise from the much lower
signal-to noise weak lensing data (and the correspondingly fewer parameters the
data can support), and the more obvious choice of basis set.

Having introduced the relevant concepts, we present in Section~\ref{sect:method}
a detailed description of the application of the atomic inference technique to
weak gravitational lensing data and demonstrate its performance on simulated
data in Section~\ref{sect:discuss}; its application to HST data is presented in
~\citet[][ to some extent]{GL/Kne++03} and in Jaunsen et al (2006, in prep.).


\section{Methodology}
\label{sect:method}


\subsection{Weak lensing background}
\label{sect:method:wl}

The data considered here are the ellipticities of $N$ background galaxies; under
the assumption of intrinsically randomly oriented galaxies, the average
ellipticity provides a (noisy) estimate of the local gravitational reduced shear
$g$ \citep[see \eg][ for an introduction]{GL/Sch06}.  The magnitude of the
(complex)  ellipticity, as used throughout this paper, is $|\epsilon| =
(1-q)/(1+q)$ where $q$ is the ellipse axis ratio. In practice, each of $2N$
lensed ellipticity components~$\epsilon_j$ (real and imaginary) are assumed  to
have been drawn independently from a Gaussian distribution with mean~$g_j$;
here~$g_j$ is the true value of the $j^{\rm th}$ component of the (complex)
reduced shear at the position of the galaxy. The likelihood function can then be
written as \citep[see \eg][ M02]{GL/SKE00}
\begin{equation}
\pr(\data|\pars) = \frac{1}{Z_{\rm L}} \exp \left(
-\frac{\chi^2}{2} \right),
\label{eq:lhood}
\end{equation} 
where 
$\data$ is the vector of ellipticity (component) values, and $\pars$ are the
parameters of the lens potential
used to calculate the shear fields at the background galaxy
positions.
$\chi^2$ is the usual misfit statistic 
\begin{equation}
\chi^2 = \sum_{i=1}^{N} \sum_{j=1}^{2} \frac{(\epsilon_{j,i} -
g_{j,i}(\pars))^2}{\sigma^2},
\label{eq:chisq}
\end{equation} 
and the normalisation factor is
\begin{equation}
Z_{\rm L} = (2 \pi \sigma^2)^{\frac{2N}{2}}.
\label{eq:chisqnorm}
\end{equation} 
The effect of errors introduced by the galaxy shape estimation procedure
have been included by adding them in quadrature to the intrinsic
ellipticity dispersion, 
\begin{equation}
\sigma = \sqrt{\sigma^2_{\rm obs}+\sigma^2_{\rm intrinsic}[1-{\rm max}(|g|^2,1/|g|^2)]^2}.
\label{eq:newsigma}
\end{equation} 
This approximation rests on the assumption that the shape estimation 
error~$\sigma^2_{\rm obs}$ is Gaussian.   Correcting for the effect of lensing
on the intrinsic ellipticity dispersion, as suggested by \citet{GL/SKE00} and
implemented by~\citet{GL/Bra++04}, effectively provides the correct weighting of
the ellipticities of images close to the critical regions of a strong lensing
cluster.  Any arclets lying within the critical curves of the cluster act as
estimators for $1/g^{*}$ instead of~$g$ \citep[\eg][]{GL/Bra++05}: we make this
correction when calculating~$\chi^2$. In practice very few galaxies are affected
by this correction, since most lie well outside the critical region, but these
are the ones that contain high lensing signal and so should be treated with
care. For a projected mass distribution $\mass(\pars)$ composed of $\Natoms$ 
physical components,  the reduced shear at position $\boldsymbol{\theta}$ is
given by
\begin{equation}
g(\boldsymbol{\theta}) = \frac{\sum_{k=1}^{\Natoms} \gamma^k(\boldsymbol{\theta})}{1 -\sum_{k=1}^{\Natoms}
\kappa^k(\boldsymbol{\theta}) },
\label{eq:redshear}
\end{equation} 
where $\gamma^k$ and $\kappa^k$ are the shear and convergence due to the $k$-th
component, \eg 
\begin{equation}
\kappa^k(\boldsymbol{\theta}) = \frac{\mass^k(\boldsymbol{\theta})}
{\mass_{{\rm crit},i}}.
\label{eq:convergence}
\end{equation} 
\citep[See, \eg,][ for more details on the weak lensing observables.]{GL/Sch06}
The (lens and source redshift-dependent) 
critical density for the $i^{\rm th}$ galaxy is $\mass_{{\rm crit},i}$ -- any
redshift information can be included here, although non-neglible redshift errors
would need to be absorbed into the likelihood function, broadening it somewhat.

Note that in the above model all mass components are assumed to be at the same
redshift. This is both necessary for the thin lens plane formulae to be
applicable, and appropriate for a cluster modelling algorithm where all mass in
the field of view is assumed to be associated with the cluster. If there were
massive clumps lying along the line of sight their gravitational shear effects
would be interpreted within this model. Extending the physical component
analysis into three dimensions is an intriguing possibility but beyond the scope
of this paper; however we do note that it may be profitable to do so, as the
additional information needed in constraining the redshift of the components,
and the more complex lens equations needed to predict the observed shear, are
both readily incorporated into the already non-linear data model.


\subsection{The physical component (halo) mass model}
\label{sect:method:atomic}

In the spherically
symmetric case the NFW density profile~\citep{CS/NFW97} is 
\begin{equation}
\rho(r) =
\frac{\rho_{\rm s}}{\left(r/r_{\rm s}\right)
\left(1+r/r_{\rm s}\right)^{2}},
\label{eq:nfwprofile}
\end{equation}
where $r_{\rm s}$ and $\rho_{\rm s}$ are the radius and density at which the
logarithmic slope breaks from $-1$ to $-3$. It is useful to normalise this
profile, which we treat as a two-parameter fitting function, to the mass
contained within a region of overdensity 200 relative to the critical density at
that redshift ~\citep{COS/All++03,COS/Evr++02}:
\begin{align}
\frac{M(r_{200})}{\frac{4}{3}\pi r_{200}^3} &= 200 \rho_{\rm crit}\\
&= 4 \pi \rho_{\rm s} r_{\rm s}^3 \left[ \log{(1+c)} - \frac{c}{1+c}\right].
\label{eq:m200}
\end{align}
Here, $c = r_{200} / r_s$ is a measure of the concentration of the halo. Lensing
properties of the NFW model have been worked out by a number of authors
\citep{GL/Bar96,GL/W+B00,GL/MBM03} and we do not reproduce their results here.
Each NFW mass component contributes to the shear field: we assign a uniform
prior distribution to the component positions within the observation region.

It has been found in many previous lensing analyses that it is rather important
to include the ellipticity of the lens~\citep[see
\eg][]{GL/KCS02,GL/San++04,GL/Koc06}. There is some choice as to whether the 
lens potential, the deflection angle or the surface density should be asserted
to be elliptically symmetric -- since the three quantities differ by the number
of times the gradient operator has been applied (0, 1 and 2 times respectively),
only one of them can have concentric elliptical contours. \citet{GL/K+S01} chose
the surface density, while \citet{GL/G+K02} opted for the slightly rectangular
projected mass  contours of a concentric elliptical deflection angle
distribution. We follow \citet{GL/MBM03} and,  citing analytic tractability, use
the lens potential~$\psi(\boldsymbol{\theta})$: the major disadvantage of this
approach is that for axis ratios of less than $\approx 0.7$ the corresponding
mass distribution becomes dumbbell-shaped.  However, massive cluster potentials
are likely to be much closer to spherical than this:  we quantify this point
below.  The breaking of axial symmetry means that the derivatives of the lens
potential  must  be calculated explicitly in order to calculate the shear and
convergence; this can be done analytically for each NFW profile component. In
this process the radius parameter is defined in such a way as to keep the mass
within a given circular radius constant as the ellipticity changes
\citep[\eg][]{GL/K+K93,GL/MBM03}.

As outlined in the introduction, N-body simulations provide the  prior
probability distributions for the NFW profile parameters.  For example,
\citet{CS/J+S02} showed that numerically simulated cluster-scale halos can be
reasonably well modelled using triaxial ellipsoidal density profiles, and
produced fitting functions for the  probability distributions of the two axis
ratios. However,  as indicated above, we prefer to use the more readily
calculated elliptical lens potential: we use the following procedure to derive
an approximate prior on the lens potential ellipticity parameter from the
distributions tabulated by \citeauthor{CS/J+S02}. 

\begin{figure}
\epsfig{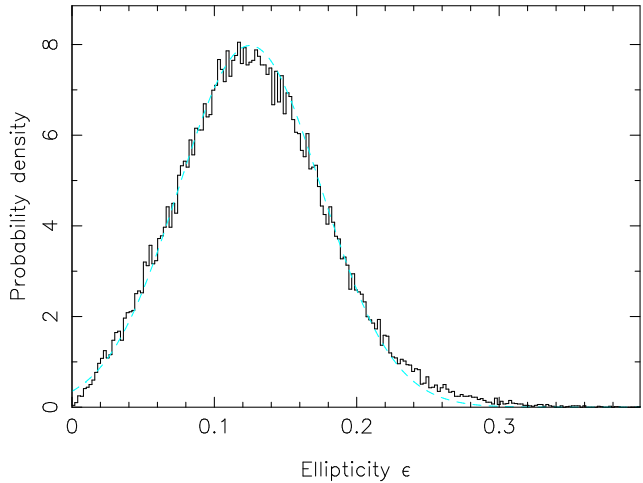}
\caption{
Prior \pdf for the (modulus of the)
lens component isopotential ellipticity, derived from fits to a 
statistical sample of
N-body simulated clusters by \citet{CS/J+S02}. The dashed curve is a 
Gaussian distribution of mean 0.125 and width 0.05. 
\label{fig:ellprior}}
\end{figure}

We draw a large number of halos' axis ratios from the fitted joint probability
density function (\pdf),  numerically project the prescribed densities onto the
lens plane, and then  compute (using, after padding with zeros, fast Fourier
transforms -- FFTs) the corresponding lens potentials via the convolution
\begin{equation}
\psi(\boldsymbol{\theta}) = \frac{1}{\pi} \int \kappa(\boldsymbol{\theta}') \log|\boldsymbol{\theta}-\boldsymbol{\theta}'| d^2\boldsymbol{\theta}'.
\label{eq:kappa2psi}
\end{equation}
We then measure the axis ratio of the resulting isopotentials, from the ratios
of the second moments of the potential. These isopotentials are not quite
elliptical, and the inferred ellipticity changes slightly with radius;  however
the derived ellipticities were compared with the actual isopotential at the
scale radius   and found to deviate by only a few percent.  The resulting
derived probability distribution for the lens ellipticity, defined as in
Section~\ref{sect:method:wl},  is given in Figure~\ref{fig:ellprior}. It is
well-approximated by a  Gaussian distribution with mean 0.125 and width
(standard deviation) 0.05,  and it is this that we use as our ellipticity
prior.  The effects of this prior are to suppress halos with unphysically high 
ellipticity, and to favour the non-spherical halos which are more commonly seen
in the  N-body simulations. The position angle of the halo is assumed to follow
a uniform distribution between $0$ and $180$ degrees: in the two-dimensional
ellipticity component space the prior \pdf peaks at the circularly symmetric
model.

The priors on the NFW profile parameters may also be derived  from N-body
simulations. For example, \citet{CS/J+S02} find the concentration parameter $c$
to be distributed log-normally with width $\approx 0.3$ about a  value of 3, 
for a massive cluster at redshift 0.5. We note that this width corresponds to an
uncertainty of $\approx 50\%$. Although much larger values of the concentration
have been inferred in some lensing analyses~\citep{GL/Kne++03,GL/Gav++03}, it is
not clear that the data have enough constraining power in the critical region to
support this conclusion. In any case, a concentration of 20 is less than 3-sigma
from the mean of the lognormal distribution, suggesting that this prior is
flexible enough to cope with unusual concentrations (whilst retaining the
desired feature of rejecting very low, and indeed negative,  values).

Finally we consider the prior on the atom mass itself, $M_{200}$. The
Press-Schecter formalism~\citep{COS/P+S74}, or one of its many
numerically-corrected forms \citep[\eg][]{COS/Jen++01,COS/Evr++02}, would serve
to provide an approximate prior \pdf for the halo mass were we observing a
random patch of sky.  The logarithmic slope of the mass function at the mass
scale of galaxy clusters is close to $-2$, indicating the rareness of massive
clusters. However, we are interested in pre-selected clusters, whose masses are
large and typically estimable to within an order of magnitude. We suggest a
compromise, and assign a Jeffreys prior \pdf (logarithmic slope $-1$) to the
halo mass, sampling uniformly in the logarithm of the mass. This has the
pleasant effect of suppressing the introduction of mass into the map unless it
is required by the data, but is not so severe that the halos either have masses
that are heavily biased low, or are discouraged completely. A more rigorous
approach would be to use the predicted halo occupation distribution to provide a
joint prior on the number and mass of sub-halos, given a main cluster component.
This is beyond the scope of this paper: the specified priors suffice to define a
robust data model. We will see in Section~\ref{sect:sim} that the method is
somewhat insensitive to the exact  form of this prior, with the presence of
components, and their masses, being principally determined from the data.


\subsection{Parameter inference}
\label{sect:method:pars}

Whilst linear methods are often favoured on the grounds of computation speed and
ease of error propagation,  we argue that the benefits of a  fully non-linear
fit outweigh these factors.  We are seeking an optimal mass reconstruction,
folding in as much information as we can: we do not wish to compromise this goal
in favour of a computationally easier alternative. With the non-linearity comes
flexibility: introducing new constraints on the mass distribution can be done in
a conceptually straightforward way, adding extra terms to the posterior \pdf for
the model parameters. 

In the non-linear data model the global best-fit point is typically hard to
find, and exists  in a parameter space whose dimensionality can run well into
double figures when many components are used. To solve these problems we employ
a Markov chain Monte Carlo (MCMC) sampler to explore the parameter space. This
technique is now widespread in astronomical data analysis  \citep[see \eg][ for
some examples of its use]
{COS/KCS01,COS/L+B02,J/MHS03,COS/Dun++05,J/Bon++04,X/PMA05}. Good introductions
to the technique are given by \citet{MCMC} and  \citet{MacKay}; here we make the
following brief comments.

Having defined a likelihood function $\pr(\data|\pars,\Natoms)$, and sensible
priors  $\pr(\pars)$ on the parameters~$\pars$ of the $\Natoms$ mass
components,  we note that the distribution containing all the information about
the mass distribution is the posterior \pdf:
\begin{equation}
\pr(\pars|\data) = \frac{\pr(\data|\pars,\Natoms)\pr(\pars)}{\pr(\data|\Natoms)}
\label{eq:bayes}
\end{equation} 
Calculating the numerator of the right hand side on a fine grid throughout the 
parameter space would allow the normalising evidence $\pr(\data|\Natoms)$ to be
calculated, the regions of high probability to be located, and any uninteresting
parameters to be numerically integrated over. In any more than a few dimensions
both these operations are  computationally unfeasible. It is much more
convenient to work with samples drawn from the posterior distribution: both
marginalisation and changing variables are trivial, the latter being done on a
sample-by-sample basis. MCMC provides an efficient way of drawing these
samples. 

Perhaps of greater practical importance is the robustness of MCMC to local
maxima in the likelihood. Optimisation schemes for locating the best-fit point
are vulnerable to becoming trapped at the wrong point in parameter space: MCMC
alleviates this problem by providing a way out of such traps (\ie with a random
trial step to a nearby part of the parameter space).

Note the dependence of the likelihood and evidence on the number of components
used in the model, $\Natoms$ (the priors on each component's parameters having
been chosen to be independent of $\Natoms$).  The posterior probability
distribution for $\Natoms$ can be seen to be available from the data via the
evidence: $\pr(\Natoms|\data) \propto \pr(\data|\Natoms)\pr(\Natoms)$. With the
assignment of a uniform prior on the number of components, the evidence gives
the (discrete) \pdf for $\Natoms$ directly.  In the same way  the evidence may
also be used to quantify the relative probabilities of two or more competing
component models, a process carried out in \citet{GL/Kne++03}. The evidence lies
at the heart of all Bayesian model selection and hypothesis testing \citep[see
\eg][ for an excellent introduction]{MacKay}, and its use is growing in
astronomy  \citep[\eg][]{COS/Jaf96,COS/Kno++98,COS/HBL02,J/MHS03,COS/MPL06}. 

We use the freely available software package \texttt{bayesys3}, written by John
Skilling. This general purpose code, used in previous work on this subject
\citep{J/MHS03}, is known to cope well with the types of likelihood surface
presented by weak lensing datasets: we find the evidence values to be accurate
and their calculation readily repeated. The evidence is calculated by
thermodynamic integration during the burn-in period \citep{Ruanaidh}.

The only disadvantage to using MCMC rather than an optimisation followed by a
Gaussian approximation to the posterior is that it can be slow: typically,
analysing a catalogue of some few thousand galaxies using a model consisting of
three components on a 3GHz processor can be expected to take several hours,
while a chi-squared minimisation may only take minutes. However, the problem of
avoiding local posterior maxima typically leads one to consider an ensemble of
minimisations from different starting points, or the use of some more advanced
algorithm such as simulated annealing (which incur many more likelihood
evaluations). Likewise, estimating uncertainties is commonly done using 
techniques such as bootstrap re-sampling or simulation of mock data.  The MCMC
process performs more or less the same calculation during the inference. 
Therefore it is the \emph{total} run time, to obtain  both reliable  parameter
estimates and their uncertainties that should be compared between methods --
with this metric MCMC becomes rather competitive.


\subsection{Probabilistic mass mapping}
\label{sect:method:maps}

While the parameters of individual halos may well be of interest
\citep[\eg][]{GL/Kne++03}, the information on the mass distribution can be
displayed in a more visually helpful way, in the form of a mass image. Each MCMC
sample corresponds to a set of halos that provide an acceptable fit to the data:
the projected mass distribution of these halos can be mapped on to a pixellised
grid. The probability distribution of the surface mass density in any given
pixel can be built up by calculating its value for each sample, and forming a
histogram. This is the \pdf marginalised over all halo parameters, so that the
width of this distribution represents the maximum uncertainty of that pixel
value, given the assumptions. To make a map, the individual pixel probability 
distributions have to be reduced to one number: we use the arithmetic mean for
its ease of calculation~\citep{ST/Ski98},  but note that an alternative central
value may be more appropriate if the pixel value \pdf is highly skewed. 

The resulting reconstructed maps inevitably retain some of the appearance of the
halos from which they are composed: however, averaging over the posterior \pdf
does bring out some extra information not present in any individual sample map.
These reconstructions can be thought of as having been highly regularised, 
using a multi-scale kernel whose shape has been chosen to be appropriate for
dark matter in cluster halos. In the next section we show some examples of these
atomic maps, and how well they describe the lensing data.


\section{Demonstration on simulated data}
\label{sect:sim}

\begin{figure*}
\begin{minipage}{0.505\linewidth}
\epsfig{file=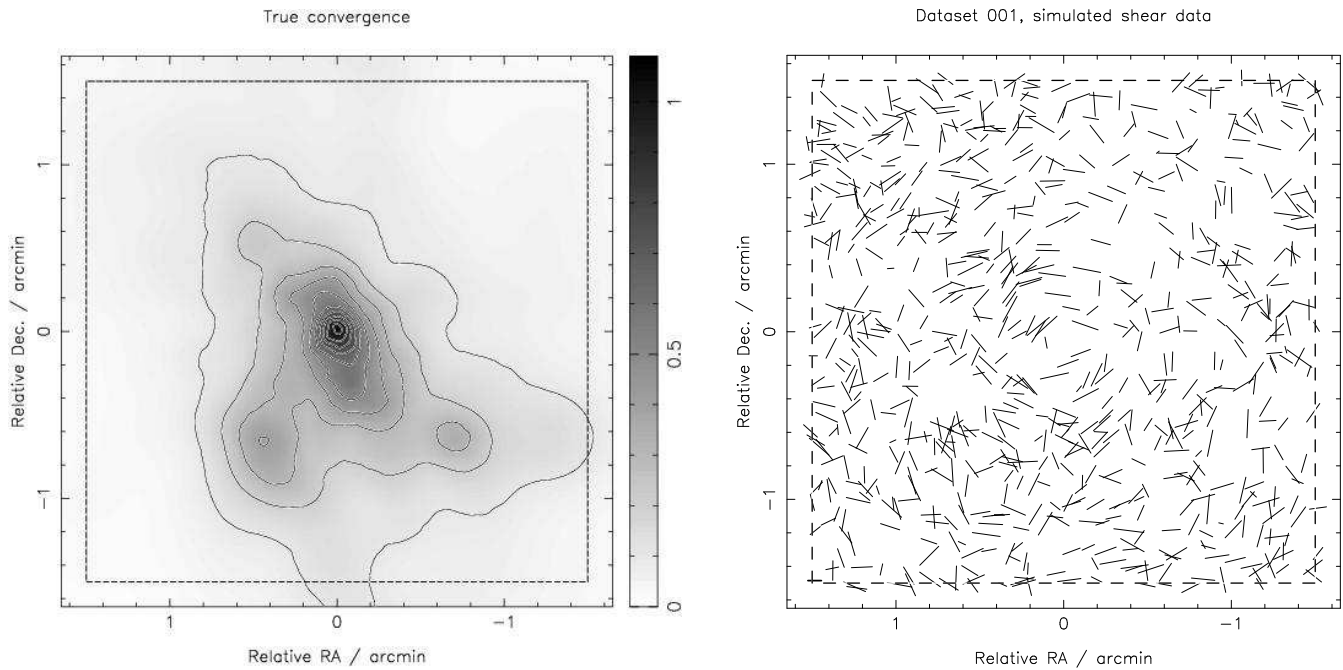,width=\linewidth,angle=0}
\end{minipage}\hfill
\begin{minipage}{0.455\linewidth}
\epsfig{file=CL09_001_shear.ps,width=\linewidth,angle=0}
\end{minipage}
\caption{
True convergence distribution (left) and mock ellipticity data (right) for the
$z=0.55$ N-body simulated cluster described in the text, assuming a source plane redshift of
1.2. The 3-arcmin square
observing region is shown by the dashed box. One main clump, and two
minor clumps, are visible in the mass map. 
\label{fig:truth}}
\end{figure*}

In order to demonstrate the methodology introduced above, we generated  mock
weak lensing data for a typical, moderately massive ($M_{\rm virial} \approx
6\times10^{14} M_{\odot}$), unrelaxed cluster at redshift~0.55. We used an
N-body simulated cluster, from the sample  in \citet*{CS/ENF98}. We placed
background galaxies at random positions on a single source plane at redshift~1.2
(such that the cluster has a tangential critical curve with radius  $\approx
5$~arcsec), with a number density of 80~per square arcmin over a square field
3~arcmin on a side. This was intended to represent a standard weak lensing
dataset from the ACS camera on HST, and resulted in a catalogue of 741 galaxy
shapes.  An intrinsic ellipticity distribution of width~0.25, and shape
estimation error of~0.2, were assumed. The reduced shear due to the input
cluster was calculated by convolving (using FFTs) the simulated cluster mass
distribution  (as provided by the simulators and after padding with zeros to
avoid edge effects) with the lensing kernel, and then scaling by the critical
density, as in \citet{GL/Mar++02}.  We note that this process used a projected 
density map with pixel scale $\approx 2$ arcsec, fine enough to retain any 
cuspy features in the mass distribution. 

When assessing the statistical performance of the atomic inference algorithm,
multiple noise realisations were used. Each realisation corresponds to a
separate set of background galaxy positions and intrinsic ellipticities, and a
different shape estimation noise term. The true projected  mass distribution,
scaled by the critical density for this lens and source redshift, is shown in
Figure~\ref{fig:truth}, along with one realisation of the simulated weak lensing
data.

A single source plane was employed for simplicity, and to separate out the
performance of the atomic modelling from systematic effects due to unknown
background galaxy redshifts. We do note again, however, that including measured
redshifts for each source is trivial in this algorithm, since we predict the
convergence and shear at each background galaxy position.


\subsection{Estimating the number of halos}
\label{sect:sim:evidence}
 
\begin{figure}
\epsfig{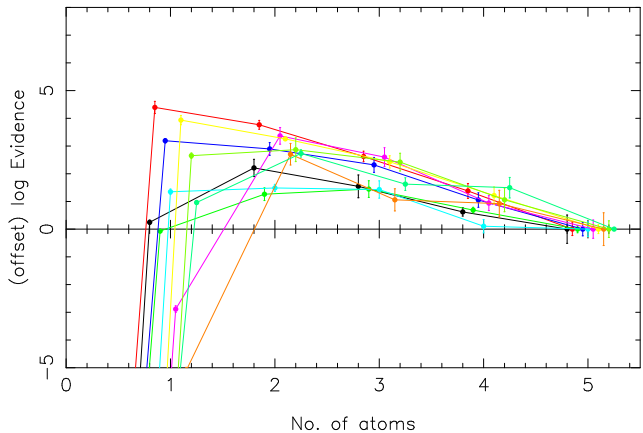}
\caption{
(Renormalised) log evidence as a function of halo number for the 
simulated cluster
described in the text. The points with error bars show the evidence estimates at each atom
number: the ten lines correspond to ten different noise realisations of the
simulation. A broad peak around atom number 1-3 is typically seen. 
\label{fig:evid}}
\end{figure}

We first investigate the number of halos appropriate for modelling this mock
dataset. This was done by sampling the posterior distributions of the parameters
of an $\Natoms$-component model, where $\Natoms$ was allowed to increase from 1
to 5.  For each inference, the evidence was calculated, and is shown in
Figure~\ref{fig:evid}. These evidence values are readily reproducible, as
indicated by the error bars on the plot: these show the standard deviations of
the mean log evidence, over 5 runs of the sampler, to be routinely less than one
unit.  This plot gives us some  confidence in the ability of the sampler to
simulate the posterior \pdf; it also quantifies the quality of the data
available in observations such as that simulated. The plot shows evidence curves
from ten different noise realisations, all of which peak at between 1 and 3
components. The different noise realisations give rise to posterior probability
distributions with different geometries and complexities: the different
challenges thus presented to the sampler result in a range of error bar sizes.
In contrast, the scatter between the curves shows the effect the noise has on
the appropriateness of $\Natoms$ components. 

The evidence for the $n$-halo model is $\pr(\data|n)$: the ratio of this to the
evidence for zero atoms (\ie the null model, with zero predicted surface  mass
density) gives a measure of the significance of the detection \citep{ST/H+M03}.
Had Figure~\ref{fig:evid} been plotted over a wider range in the ordinate, 
probability (evidence) ratios relative to the null model of ${\rm e}^{40-60}$
would have been visible, indicating a resounding detection of the cluster in the
shear data. The differences between evidence values for $n$ of 1, 2, 3, 4 and 5
components  are much smaller, typically reaching between two and four units in
the logarithm between the peak evidence and the $n=5$ value. These differences
correspond to probability ratios of approximately $10-50$: the data are found to
be typically an order of magnitude more likely to have come from a two-component
model than a 5-component model (with the exact favoured value of $n$  being
dependent on the noise realisation). This is an important result: the simulated
data were designed to be representative of that being analysed in contemporary
work; what we are seeing here is a quantification of the amount of information
available to us from that data. With the well-appointed halo model used, only a
handful of parameters are both required and supported by the data. This is in
agreement with the findings of \citeLE: in the next section we show how the
inferred mass distributions differ from the maximum entropy maps. 

\subsection{Mass mapping}
\label{sect:sim:maps}

\begin{figure*}
\begin{minipage}{0.32\linewidth}
\epsfig{file=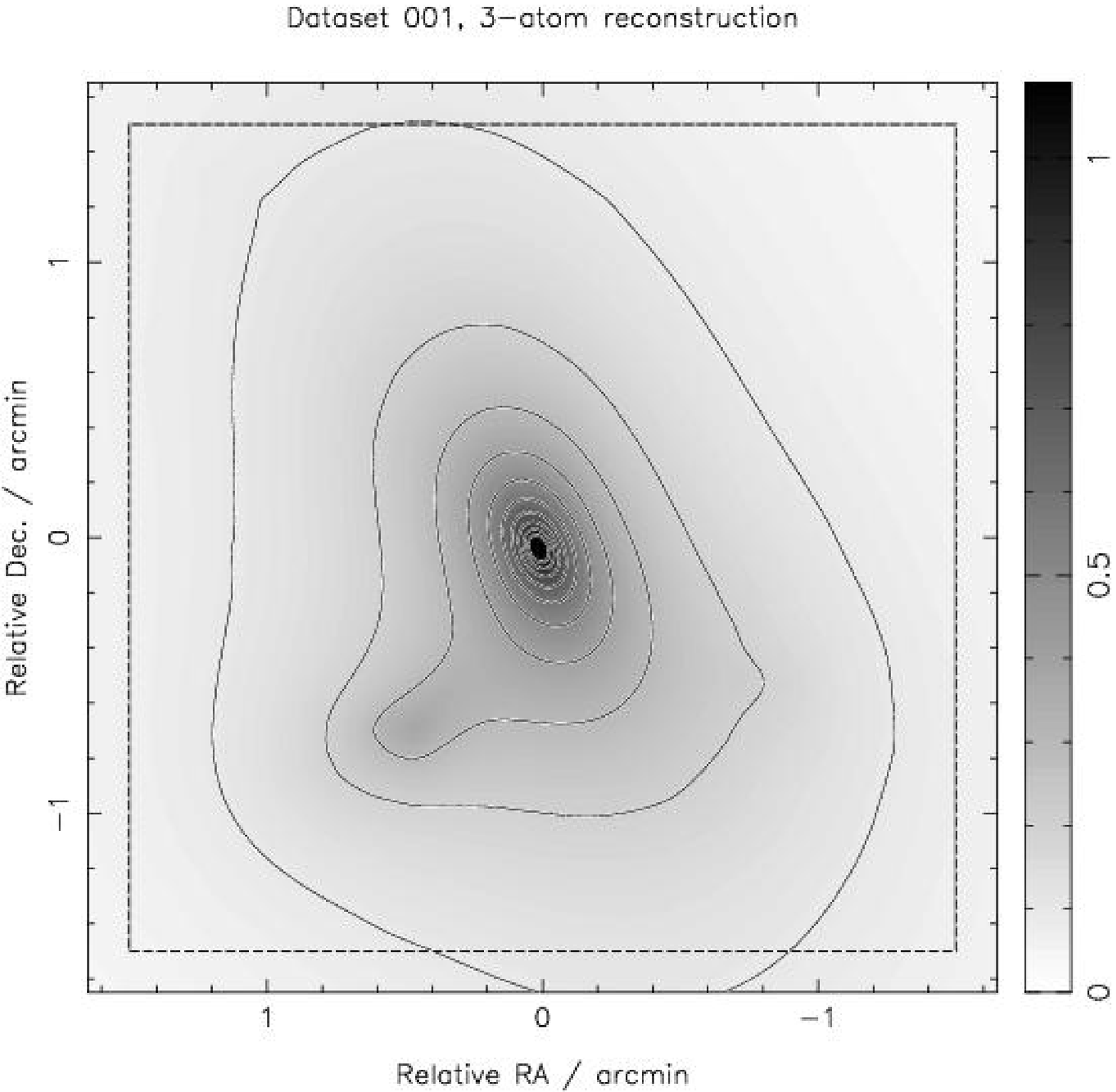,width=\linewidth,angle=0}
\end{minipage}\hfill
\begin{minipage}{0.32\linewidth}
\epsfig{file=CL09_001_LE_maxent_015.ps,width=\linewidth,angle=0}
\end{minipage}\hfill
\begin{minipage}{0.32\linewidth}
\epsfig{file=CL09_001_LE_maxent_040.ps,width=\linewidth,angle=0}
\end{minipage}

\begin{minipage}{0.32\linewidth}
\epsfig{file=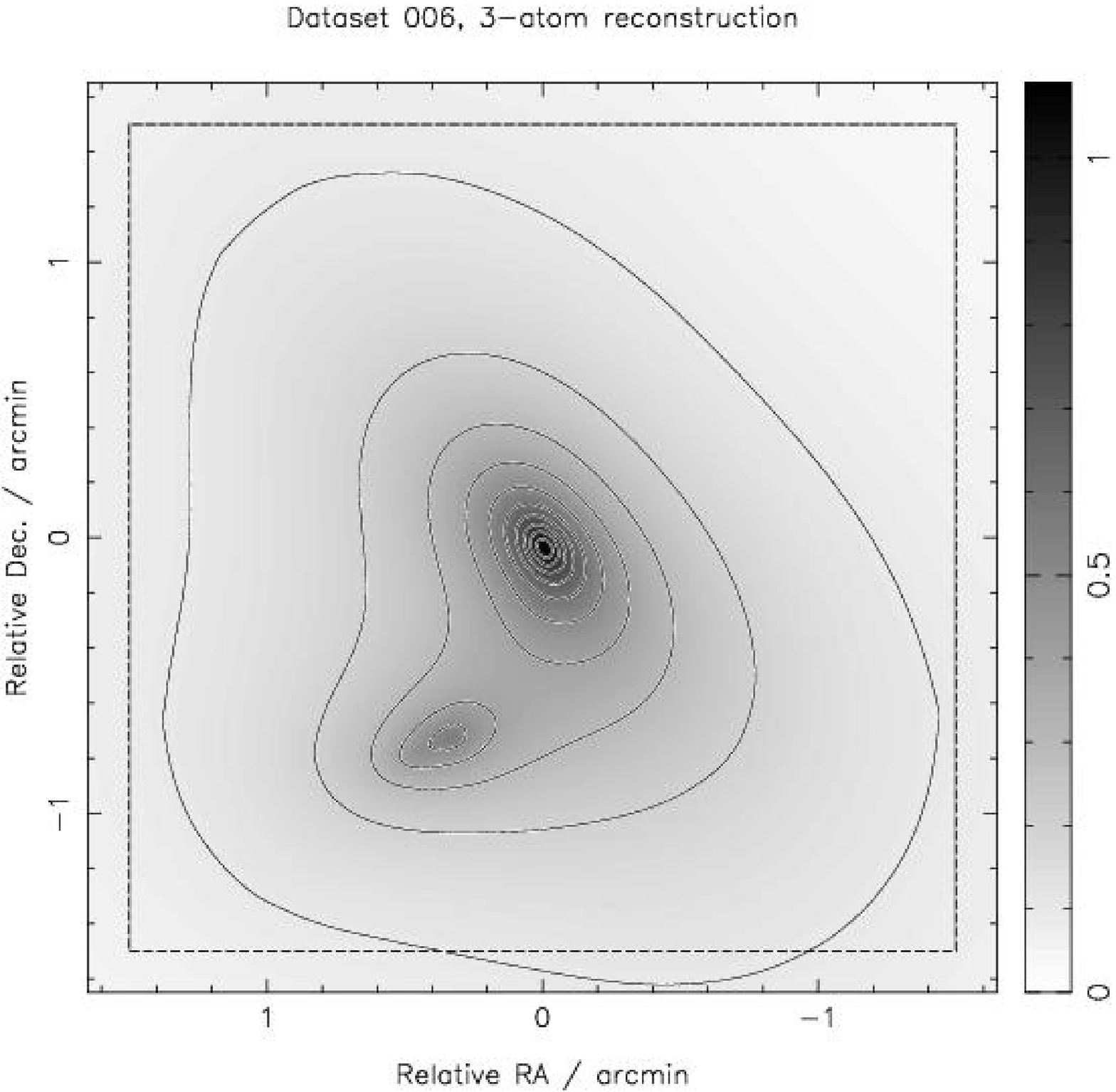,width=\linewidth,angle=0}
\end{minipage}\hfill
\begin{minipage}{0.32\linewidth}
\epsfig{file=CL09_006_LE_maxent_015.ps,width=\linewidth,angle=0}
\end{minipage}\hfill
\begin{minipage}{0.32\linewidth}
\epsfig{file=CL09_006_LE_maxent_040.ps,width=\linewidth,angle=0}
\end{minipage}
\caption{
Two representative mass reconstructions, one noise realisation per row,
for the
$z=0.55$ simulated cluster described in the text. 
Left column: ensemble-average atomic inference mass distributions.
Centre and right columns: 15 and 40-arcsec resolution maximum-entropy maps.
These panels 
may
be compared directly with the left-hand panel of Figure~\ref{fig:truth}.
The 3-arcminute square
observing region is shown by the dashed box. 
}
\label{fig:maps}
\end{figure*}

The left-hand panels of Figure~\ref{fig:maps} show the ensemble-averaged
halo-model mass reconstructions, for two different noise realisations. For
comparison, the maximum-entropy maps are shown along side, at two different
resolution scales. These plots serve to illustrate some general points about the
two techniques. The \LE algorithm, and any other method that uses a single
resolution scale when smoothing the data or in the reconstruction process
itself, does not cope well with the range of scales of mass structure in this
cluster. A sharp, cuspy peak, surrounded by an extended irregular mass
distribution in the outer regions clearly requires at least two scales
(approximated here by the 15 and 40-arcsec \LE resolution kernels): these scales
are provided naturally by the physical components. The ``atomic'' reconstruction
is remarkably robust between noise realisations, whereas the more flexible
pixel-based method maps contain transient structure as the noise is fitted. This
problem was alleviated in \citeLE by increasing the resolution scale until, at
maximum evidence, only believable features remained; the same may be said about
the structure in the atomic maps, except that the small-scale, high
signal-to-noise structure is retained (\eg at the centre of the cluster).
Finally we note the long standing problem of inferring super-critical density
from weak shear data, outlined in some detail in the work of
\citeauthor{GL/S+S95}~(\citeyear{GL/S+S95,GL/S+S95b}). With no additional
information it is impossible to infer uniquely the presence of convergence
greater than unity: in contrast, when constructing the mass distribution from
naturally cuspy components the convergence can be effectively interpolated
upwards in a  seamless and physically-reasonable way.

\begin{figure}
\begin{minipage}{\linewidth}
\epsfig{file=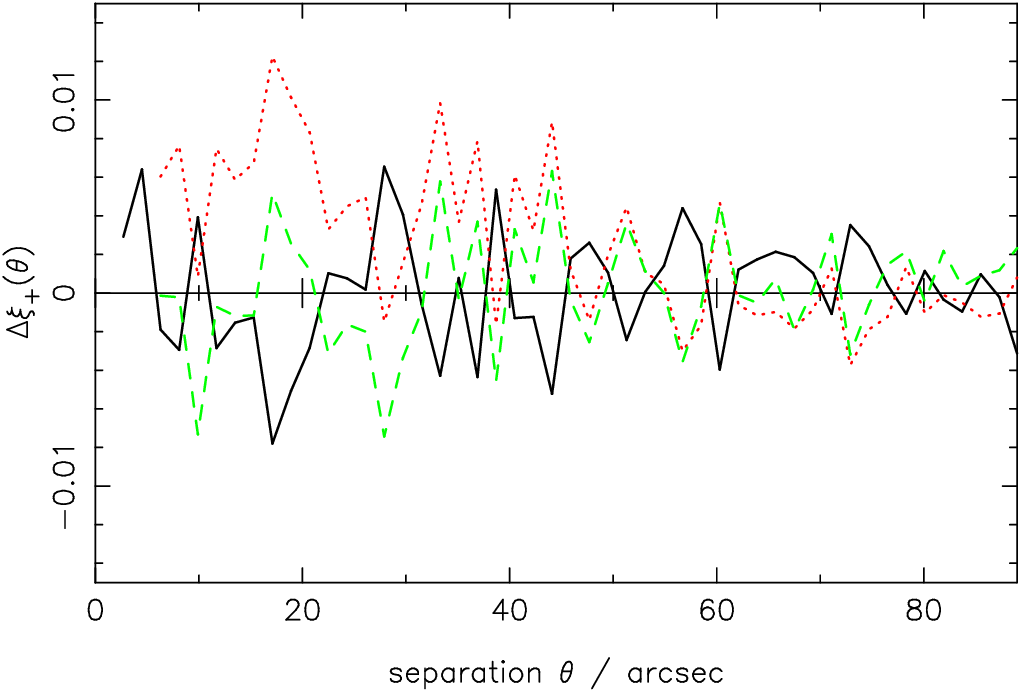,width=\linewidth,angle=0}
\end{minipage}
\begin{minipage}{\linewidth}
\epsfig{file=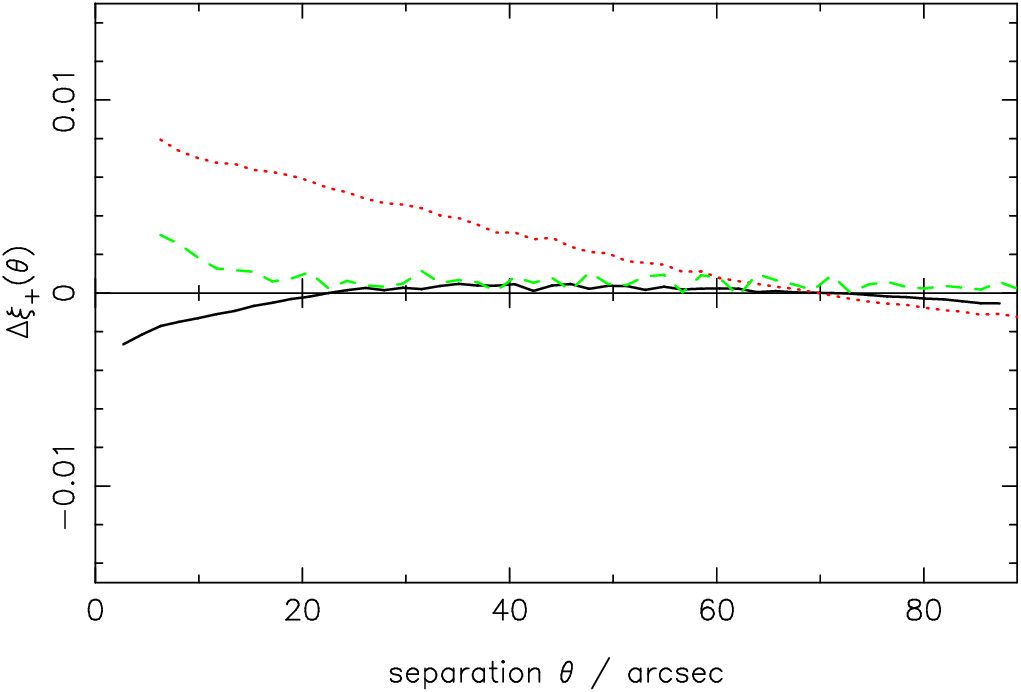,width=\linewidth,angle=0}
\end{minipage}
\caption{
Quantifying the mass map reconstruction accuracy via the correlation function
differences.
Top: correlation function~$\xi_{+}$ of the difference between the 
ensemble-average 
predicted reduced shear, and the data. The curves shown are for 
the 3-halo atomic reconstruction (solid), 15-arcsec resolution \LE map 
(dotted), and the
40-arcsec  \LE map (dashed).
Bottom: the same exercise, with the same legend, 
but now comparing the predicted reduced shear with the true
input values. 
\label{fig:corr}}
\end{figure}

The observations of the previous paragraph can be put on a more quantitative
footing by plotting the correlation between the inputs and outputs of the
algorithms. This is shown in Figure~\ref{fig:corr}. We use the correlation
function $\xi_{+}(\theta)$, where $\theta$ is the angular separation between
galaxy pairs;  this is given by \citep[\eg][]{GL/Sch06}:
\begin{equation}
\xi_{+}(\theta) = \langle g_{\rm t}^{\rm A} g_{\rm t}^{\rm B}\rangle
                + \langle g_{\rm x}^{\rm A} g_{\rm x}^{\rm B}\rangle,
\end{equation}
where $g_{\rm t}^{\rm A}$ ($g_{\rm x}^{\rm A}$) is the tangential (cross)
component  of the reduced shear estimator at galaxy position of galaxy A,
relative to galaxy position B (and vice versa). This function conveniently
quantifies the alignment of pairs of galaxy shapes.  Since we are interested in
the difference between the  reduced shear predicted by the reconstructions
($g$), and either the true reduced shear ($\hat{g}$) or measured ellipticities
($\epsilon$), we construct the difference function
\begin{align}
\Delta \xi_{+}(\theta) &= \langle (g_{\rm t}^{\rm A} - \hat{g}_{\rm t}^{\rm A})
                                 (g_{\rm t}^{\rm B} - \hat{g}_{\rm t}^{\rm
                                 B})\rangle \notag \\
                       &\;\;\;\;+ \langle (g_{\rm x}^{\rm A} - \hat{g}_{\rm x}^{\rm A})
                                 (g_{\rm x}^{\rm B} - \hat{g}_{\rm x}^{\rm B})\rangle \\
                       &= \xi_{+}^{gg}(\theta) + \xi_{+}^{\hat{g}\hat{g}}(\theta) 
                            - 2 \xi_{+}^{g\hat{g}}(\theta).
\end{align}
For a perfect match on all scales, this  function would be zero.  All
correlation functions decrease with increasing pair separation, as the lensing
signal diminishes in strength. The upper panel of Figure~\ref{fig:corr} shows
that  the residuals in the 3-atom reconstruction, and the evidence-preferred
40-arcsec  \LE reconstruction, are consistent with noise; the high resolution
\LE reconstruction shows a small positive  $\Delta \xi_{+}(\theta)$ at scales of
5-40 arcsec indicative of an imperfect reconstruction.   This shortcoming is
seen more clearly in the lower panel; in the figure the low resolution \LE map
and the 3-atom reconstruction are seen to perform roughly equally well in
recovering the true mass distribution, with the atomic map doing slightly better
on the smallest scales. This agrees with the maps of Figure~\ref{fig:maps},
where the cuspy cluster centre is not reproduced with the smooth maximum-entropy
map.

\begin{figure*}
\begin{minipage}{0.32\linewidth}
\epsfig{file=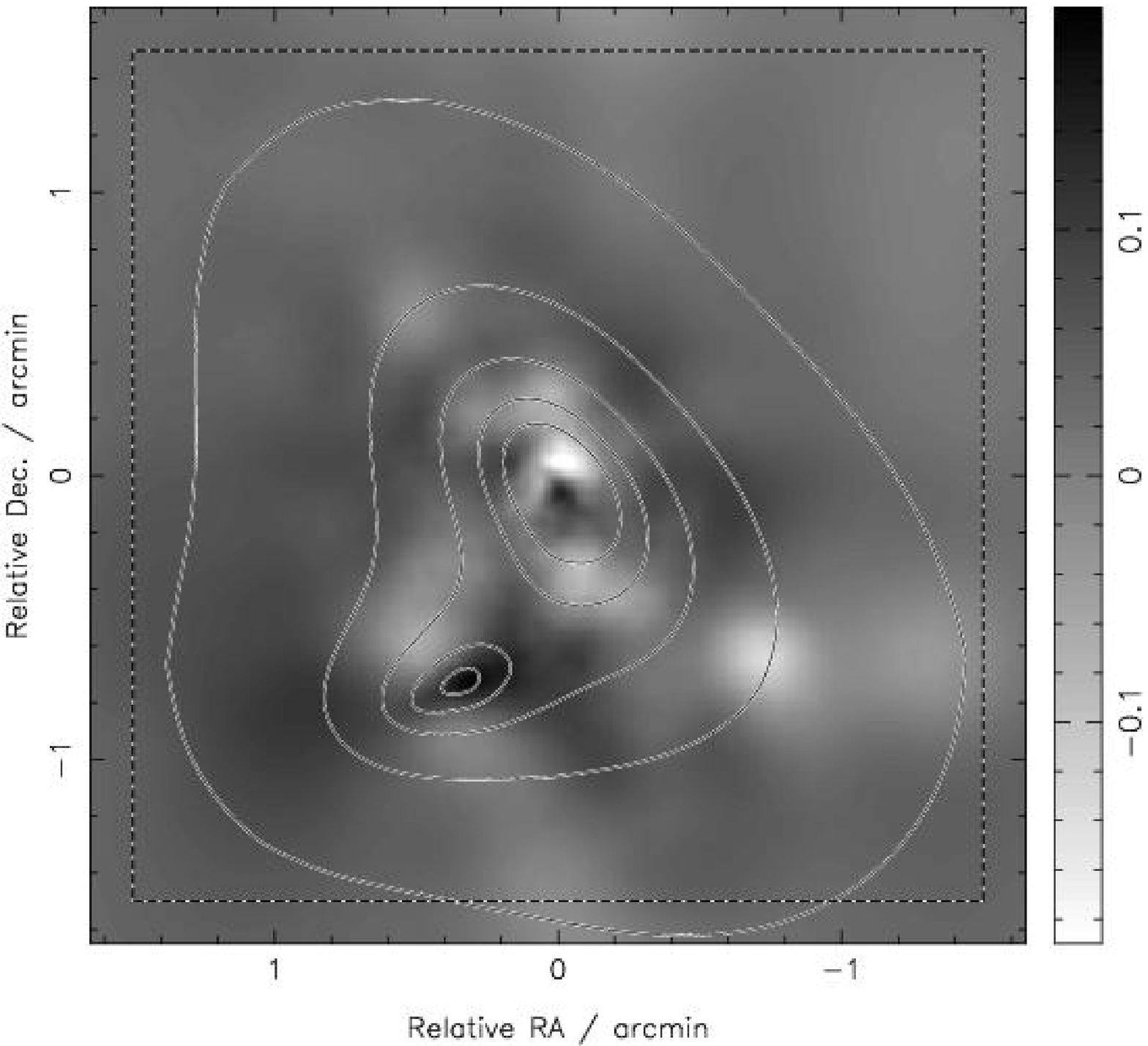,width=\linewidth,angle=0}
\end{minipage}\hfill
\begin{minipage}{0.32\linewidth}
\epsfig{file=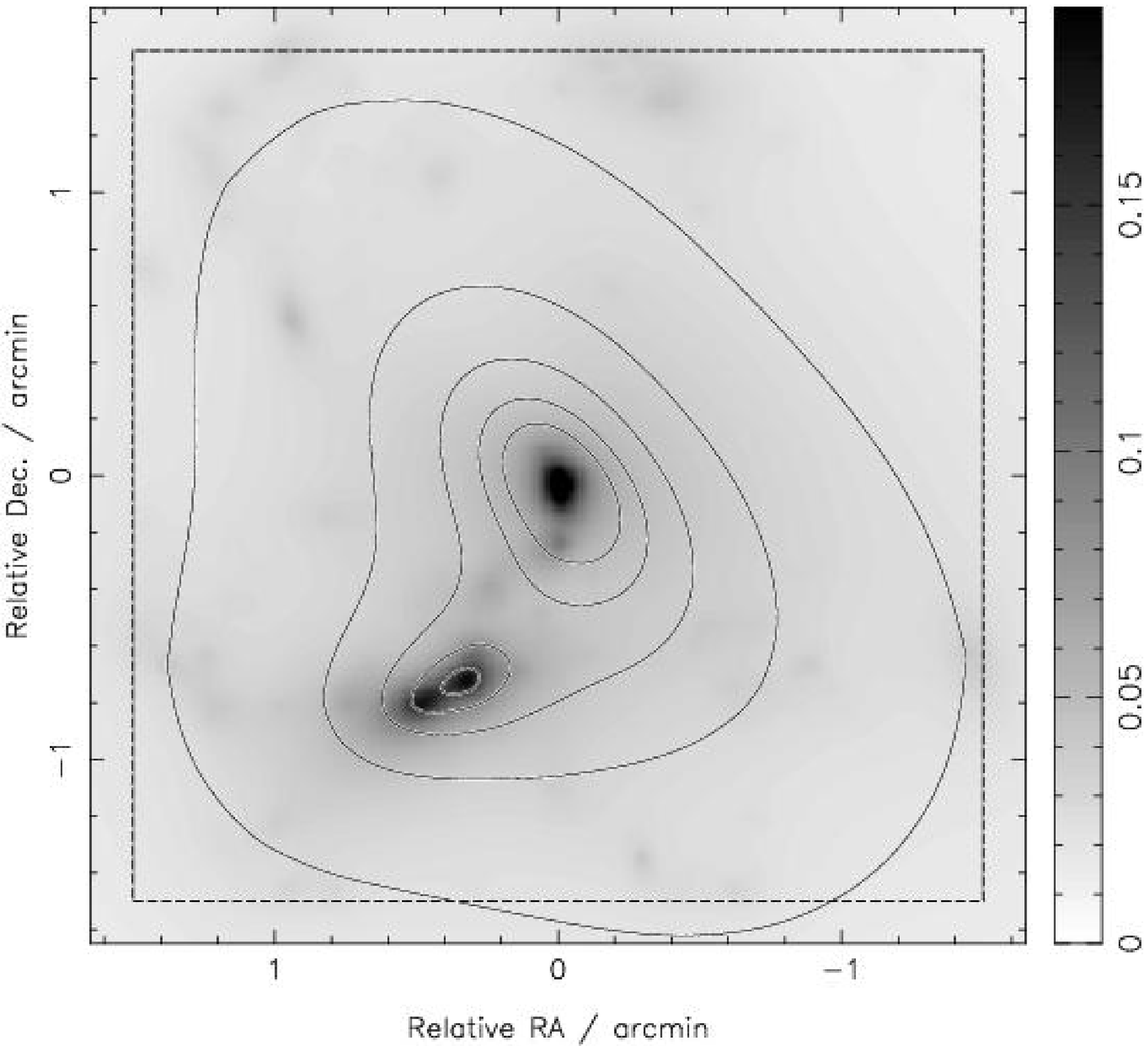,width=\linewidth,angle=0}
\end{minipage}\hfill
\begin{minipage}{0.32\linewidth}
\epsfig{file=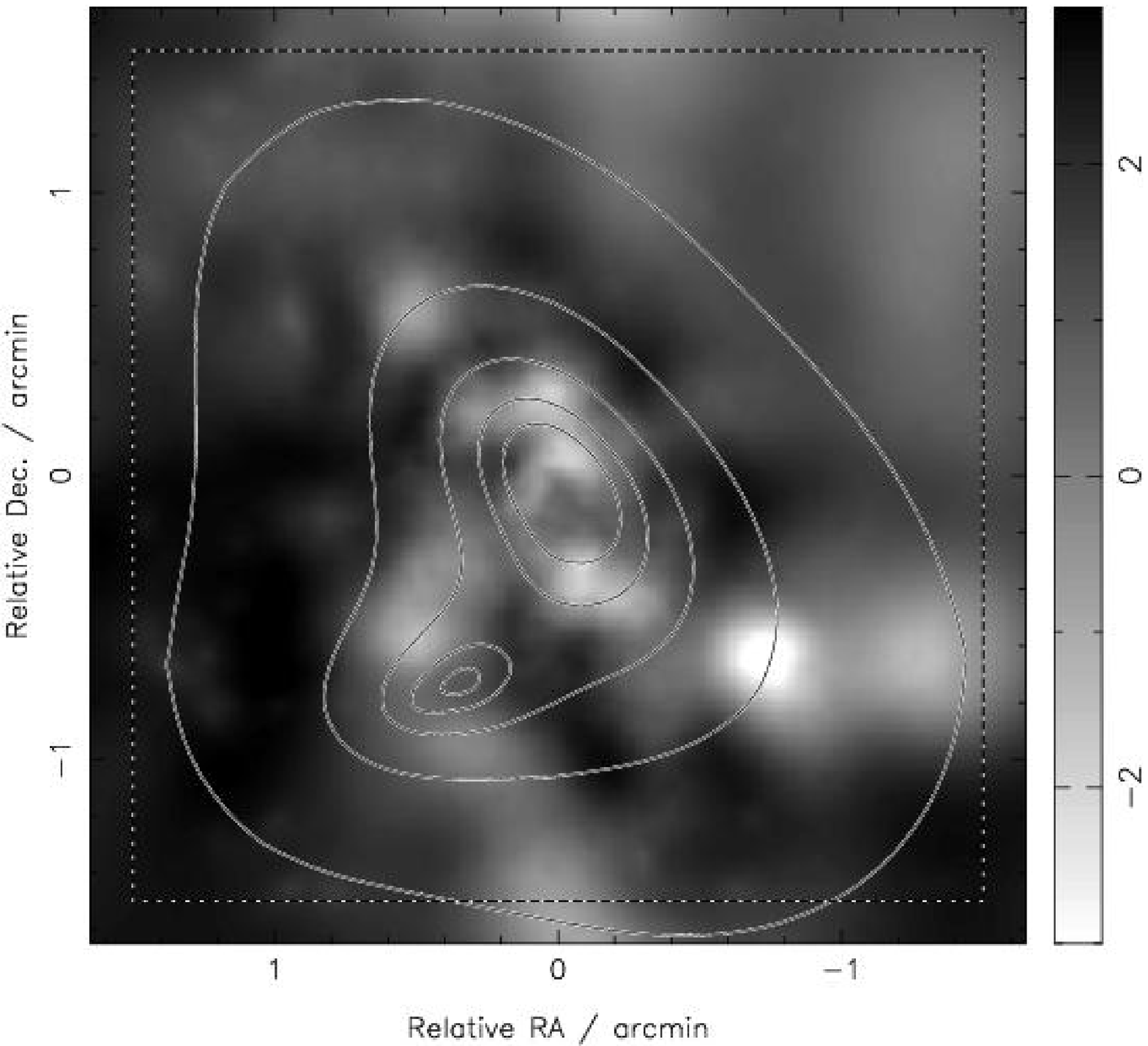,width=\linewidth,angle=0}
\end{minipage}
\caption{
Left: difference map between the 3-halo reconstruction of Figure~\ref{fig:maps} 
and the
true surface density seen in Figure~\ref{fig:truth}.
Centre: the physical component method 
uncertainty map, as estimated by the standard deviations of the
individual pixel {\pdf}s. 
Right: the difference map, divided by the error map.  
In all panels the reconstructed convergence contours are overplotted to guide
the eye.
\label{fig:error}}
\end{figure*}

The information in the correlation functions may also be visualised via  maps of
residuals. Plotted in the left panel of  Figure~\ref{fig:error} is the
difference  between the  reconstruction in the top left hand panel of
Figure~\ref{fig:maps}, and the true mass distribution; for comparison, the
central panel shows the width of the pixel value \pdf, as estimated by the
standard deviation of the samples. This ``error''  map is informative: even in
the regions where the shear signal is strong, the uncertainty on the predicted
convergence is high. However, using this uncertainty map to rescale the
residuals between reconstruction and truth we see that the differences are
fairly low significance: only in the region of the  undetected sub-clump are the
pixel values more than 3 sigma from the truth (where ``sigma''  is the
uncertainty mapped in the central panel). 

Evident in Figure~\ref{fig:error} is the small-scale mass structure in the
cluster core, undetectable by weak lensing, that makes pointwise convergence
prediction at the level of a few percent or better very difficult. Clearly more
information is needed: the work of \citet{GL/N+S04} indicates that including the
mass associated with galaxies (via small components placed at the cluster
member  positions) is sufficient  to model this substructure. 

\subsection{Component properties}
\label{sect:sim:components}

\begin{figure}
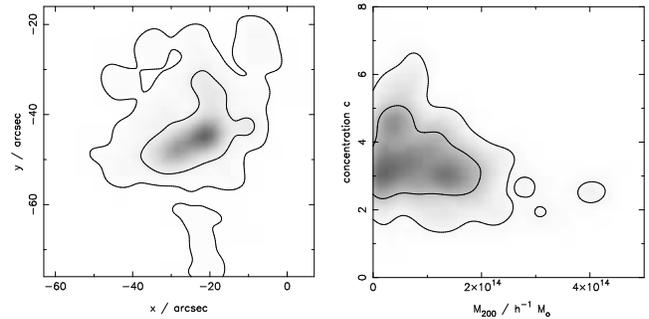

\begin{minipage}{0.48\linewidth}
\epsfig{file=CL09_newellprior+errcorr_006_3_pars.subclump_x-y.ps,width=\linewidth,angle=0}
\end{minipage}\hfill
\begin{minipage}{0.48\linewidth}
\epsfig{file=CL09_newellprior+errcorr_006_3_pars.subclump_M200-c.ps,width=\linewidth,angle=0}
\end{minipage}
\caption{
Inferences about the south-eastern substructure seen in the reconstruction of
Figure~\ref{fig:maps}. Left: $\pr(x,y|\data)$; right: $\pr(M_{200},c|\data)$.
In the right panel the prior on the concentration can be seen, allowing a
reasonable estimate of the mass of the marginally-detected subclump. 
\label{fig:subclump}}
\end{figure}

The ensemble average mass map is one way of representing the information in the 
joint posterior \pdf; marginal distributions for other parameters of interest
are also readily available. For example, Figure~\ref{fig:subclump} shows the
position, and  mass profile parameters, associated with the secondary sub-clump
visible in the projected mass map. MCMC samples corresponding to a circular
region centred on the map peak were excised and histogrammed, to plot the
marginalised posterior distributions for component position $\pr(x,y|\data,H)$,
and density profile parameters
$\pr(M_{200},c|\data,H)$, where $H$ is the assumption that a mass feature of
interest lies within this aperture. The widths of these distributions provide
estimates of the uncertainties on the parameters. This secondary feature is
somewhat transient: in some noise  realisations the signal is too
broken up to be detected. 

\section{Discussion}
\label{sect:discuss}

We now discuss three of aspects of the physical component analysis in greater
detail. First is the sensitivity of the method to the prior {\pdf}s assigned to
the model parameters;  second is the effect of the  assumed halo model on the
accuracy of the reconstruction; third  is the potential for including strong
lensing information.

\subsection{``Bias'' in the reconstruction}

We point out explicitly that the mass maps derived in the physical component
analysis are biased, in the sense that no matter how good the data is the
predicted mass distribution still has to be constructed  from NFW-shaped halos,
and in any individual cluster observed at a high signal-to-noise ratio one can
imagine this model breaking down. Indeed, careful inspection of the maps in
Figure~\ref{fig:error} shows that the reconstructed  surface density is
systematically higher than the true density in the outer parts of the field of
view, by 0.05 in convergence or so. This is despite the ellipticity data being
fitted to a fully satisfactory level (Figure~\ref{fig:corr}), and is due to  the
mass sheet degeneracy ~\citep{GL/FGS85}. This effect when working with
parameterised profiles was pointed out by  \citet{GL/SKE00} and further
investigated by \citet{GL/Bra++04},  and was left as a limitation on the
measurement of cluster density profiles. In the present case  we are effectively
selecting a particular mass sheet transformation parameter through our choice of
the NFW profile for our mass components: the error bars mapped in
Figure~\ref{fig:error} are model-dependent. The bias does lie, however, within
this statistical uncertainty: the simulated cluster used here is well-fit by a
density profile close to NFW! 

The model-dependence of the inferences advocated is therefore presented as a
sensible use of the prior information available from N-body simulations.
However, one should remember that the universal profiles came from fitting the
ensemble average profiles of simulated halos: in any given cluster it may be
that the prior distribution for the concentration may not afford the halo
profiles sufficient freedom to fit the data well, or even that some alternative
profile is more appropriate altogether. In this case, the profile can be
inferred from the data via the evidence as was shown in \citep{GL/Kne++03}. When
using any other profile, the priors on the halo  parameters will not be as
readily derived, and indeed may be preferred to be kept uninformative. In this
case the Occam's razor factor inherent to the evidence will act to favour the
basic (and \emph{a priori} better-constrained) NFW atom set. A consequent
increase in evidence when using the alternative profile will then be a rather
robust conclusion, the analyst's natural tendency towards the expected forms
having been taken care of already.

\subsection{Robustness to prior {\pdf}s}
 
Weak lensing data is very noisy: in such situations, we should be wary of  the
prior \pdf  becoming comparable in importance to the likelihood function.  We
investigated this for the simulated dataset described above by applying some
alternative priors and examining the evidence values. For the component masses
we tried a prior uniform in mass (as opposed to the Jeffrey's prior suggested
above, which is uniform in the logarithm of the mass). This prior is more
favourable to higher mass components: if the presence of an extra component were
being suppressed by the Jeffreys prior then we would be more likely to see it
with the more forgiving uniform distribution. We found the Jeffrey's prior to
give slightly higher evidence, perhaps slightly better reflecting the true halo
occupation distribution.  The difference was however, much less than one unit,
suggesting that the component number and mass are well-determined by the data.
Indeed, the resulting mass maps were indistinguishable from those from the main
analysis. 

Connected to the prior for the component mass is that for the component number,
implicitly assumed to be uniform when interpreting Figure~\ref{fig:evid}. While
there may be some theoretical motivation for attempting to model the cluster
with a mass function, that gives a joint \pdf on $\Natoms$ and halo mass, that
is steeper than that used here, we would run the risk of over-fitting the data.
At present,  Figure~\ref{fig:evid} shows that the fit is being sensibly
regularised in models with small numbers of components,  with the evidence
clearly decreasing towards higher~$\Natoms$. When using data with artificially
low noise, the evidence does indeed favour higher numbers of components, and
lower mass features in the field are indeed recovered, as expected. 

We also investigated the effect of changing the component ellipticity prior,
broadening it to make higher ellipticity components intrinsically more probable.
Again, this made very little difference to the evidence values or to the mass
maps.

\subsection{Including strong lensing information}

The non-linear nature of the inference process described in the previous
sections makes it straightforward to include strong lensing constraints as
priors.  These constraints may come from galaxy-scale lenses in the field, or
from objects multiply-imaged by the cluster potential itself.

For example, an independent strong lens modelling (on either scale)  might be
designed to yield estimates of the convergence and shear (or perhaps more
likely, magnification $\mu$) at a particular point, with error bars
(\eg~$\sigma_{\mu}$). These uncertainties, when translated into Gaussian (or the
appropriate error distribution) {\pdf}s centred on the point estimate, can be
used to weight the MCMC trial parameter sets: the weak lensing likelihood is
simply multiplied by the value of a \pdf such as $\pr(\kappa|\mu,\sigma_{\mu})$.
This approach, while convenient for marrying two independent pieces of analysis
software, would be somewhat wasteful in its use of information (and has not yet
been applied in practice!). A more rigorous approach would be to include the
strong lens image positions, or even the image pixel values themselves, as data,
and form a joint likelihood from the product of the individual weak and strong
lensing forms.  Such an analysis is beyond the scope of this work, but is being
developed as an extension of the {\scshape LensTool} package \citep{GL/Kne++96}.
MCMC is anticipated as being very useful indeed here: the strong lensing
likelihood surface is highly complex.

One application of cluster lensing that has been suggested is to predict the
``external'' convergence at the position of galaxy-scale strong lenses in the
cluster field \citep[\eg][]{GL/B+F99,GL/KCH00}.  Any contribution to the strong
lens convergence is degenerate with the estimate of Hubble's constant from that
lens, motivating the study of additional constraints on the convergence, such as
might be hoped for from weak lensing. While both strong and weak lens systems
suffer from the mass sheet degeneracy referred to above, when assuming
physically-motivated parameterised profiles for all mass distributions this
degeneracy is broken. A model-dependent  estimate of convergence at a particular
point in the field (and indeed its posterior probability distribution) can then
be generated from the MCMC samples: its accuracy will be dependent on both the
noise in the weak lensing data, and the assumptions of an NFW-profile  component
cluster. The maps in Figure~\ref{fig:error} suggest that, for the  data quality
investigated here, both the statistical and systematic parts  of the error bar
on a pointwise convergence estimate are likely to be in the region of 0.05,
representing a significantly greater fractional error than that due to the
best-measured time delays.


\section{Conclusions}
\label{sect:concl}

We have presented a new method for reconstructing the mass distribution of
clusters of galaxies from weak gravitational lensing data, based on the atomic
inference procedure suggested by \citet{ST/Ski98}. By investigating its
performance on simulated HST ACS data, we draw the following conclusions:

\begin{itemize}

\item Perhaps as expected, the number of mass components supported by the data
(and selected via the evidence) is typically quite small. This is a consequence
of the domination of clusters by a single deep potential with a few satellite
halos, combined with the paucity of information on the weak shear data. 

\item The physical component (posterior average) mass maps are cleaner and more
robust to the noise realisation, than those made with pixel-based methods: the
natural basis set of elliptical NFW profile halos act to suppress  spurious
peaks and provide an accurate reconstruction. This accuracy is evident in the
high-strength  correlation between input map and reconstruction, that extends
(by virtue of the additional information input to the model) to smaller angular
scales than previous techniques. 

\item The mass maps are biased towards the results of numerical simulations,
incorporating our expectations of what clusters should look like. As a result,  
the mass sheet degeneracy is broken, and absorbed into the uncertainties
associated with the maps. The accuracy of the reconstruction is
partly determined by the assumption of the NFW profile for each halo, but this
is an assumption that can be ranked against competing profiles using the
evidence.

\item The necessarily non-linear method, while slow to execute, has the
advantages of coping with local minima in the chi-squared surface, providing
error bars on the model parameters directly, and easily accommodating additional
physical constraints of an arbitrary functional form.

\end{itemize}

This very last point opens the door for a more comprehensive weak plus strong
lensing reconstruction strategy:  the method outlined here is conceptually very
clean, and consequently provides a framework within  which  additional
constraints on the cluster mass distribution can be straightforwardly
incorporated. 

The (standard fortran77 and c) code used in this and the cited work is available
on request from the author.


\section*{Acknowledgments}

We thank Mike Hobson and Steve Gull for useful discussions on atomic inference
methods, and John Skilling for moreover providing his BayeSys3 code (freely
available from \texttt{http://www.inference.phy.cam.ac.uk/bayesys/}). We also
thank Jean-Paul Kneib, Roger Blandford for their comments and advice, Patrick
Hudelot and Farhan Feroz for helping test the {\scshape McAdam} code,  and
Maru{\v s}a Brada{\v c} for a careful reading of the manuscript. We are grateful for
the encouragement and constructive criticism of the referee,  Peter Schneider,
that led to a much-improved piece of work. This work was supported in part by
the U.S. Department of Energy under contract number DE-AC02-76SF00515.


\label{lastpage}
\bibliographystyle{mn2e}


\bsp

\end{document}